\documentclass[twocolumn,aps,prb,longbibliography,superscriptaddress,floatfix]{revtex4-2}

\usepackage[T1]{fontenc}
\usepackage[utf8]{inputenc}
\usepackage[english]{babel}
\usepackage{color}
\usepackage{float}
\usepackage{braket}
\usepackage{amsmath}
\usepackage{amstext}
\usepackage{amssymb}
\usepackage{graphicx}
\usepackage{bbold}
\usepackage[unicode=true,pdfusetitle,
bookmarks=true,bookmarksnumbered=false,bookmarksopen=false,
breaklinks=false,pdfborder={0 0 1},backref=false,colorlinks=false]
{hyperref}
\newcommand{\angstrom}{\textup{\AA}} 
\hypersetup{
	colorlinks,linkcolor=blue,citecolor=blue,urlcolor=blue}
\usepackage{soul}
\makeatletter
\date{}

\makeatother
\setul{0.5ex}{0.3ex}
\definecolor{Blue}{rgb}{0,0.0,1}
\setulcolor{Blue}
\usepackage{babel}

\begin{document} 

\author{Tarik P. Cysne}
\email{tarik.cysne@gmail.com}
\affiliation{Instituto de F\'\i sica, Universidade Federal Fluminense, 24210-346 Niter\'oi RJ, Brazil} 

\author{Ivo Souza}
\affiliation{Centro de F\'isica de Materiales, Universidad del Pa\'is Vasco, 20018 San Sebast\'ian, Spain} 
\affiliation{Ikerbasque Foundation, 48013 Bilbao, Spain} 

\author{Tatiana G. Rappoport}
\affiliation{Centro Brasileiro de Pesquisas F\'isicas , R. Dr. Xavier Sigaud, 150, 4715-330 Rio de Janeiro, RJ, Brazil} 
\affiliation{Physics Center of Minho and Porto Universities (CF-UM-UP),Campus of Gualtar, 4710-057, Braga, Portugal} 
\affiliation{International Iberian Nanotechnology Laboratory (INL), Av. Mestre Jos\'e Veiga, 4715-330 Braga, Portugal}

\title{Orbital Hall effect from orbital magnetic moments of Bloch states: the role of a new correction term}

\begin{abstract}
We present a rigorous derivation of the matrix elements of the orbital magnetic moment (OMM) of Bloch states. Our calculations include the Berry connection term in the k-derivatives of Bloch states, which was omitted in previous works. The resulting formula for the OMM matrix elements applies to any non-degenerate Bloch states within Hilbert space. We identify two new contributions: the first restores gauge covariance for non-degenerate states, while the second, being itself gauge covariant, can provide significant quantitative corrections depending on the system under study. We examine their impact on the orbital Hall effect in two bilayer systems: a 2H transition metal dichalcogenide bilayer and a biased bilayer graphene. In both cases, these new terms reduce the orbital Hall conductivity plateau compared with results that neglect them, suggesting that multi-layered van der Waals materials may be particularly susceptible to the derived OMM corrections. Our findings may contribute to the formal understanding of electronic OMM transport and to the conceptual foundations of the emerging field of orbitronics.  
\end{abstract} 
\maketitle 

\section{Introduction}

In recent years, the physics of orbital angular momentum (OAM) associated with various excitations in solids has acquired increasing significance \cite{Mag-OAM-doi:10.1021/acs.nanolett.4c00430, Phonon-OAM-doi:10.1021/acs.nanolett.0c03220, Magnons-PhysRevLett.125.117209, Mag-OAM-PhysRevLett.129.167202}. In particular, electronic OAM has revealed a wide range of previously unexplored phenomena, such as the orbital Hall effect (OHE) \cite{Bernevig-PhysRevLett.95.066601, Go-PhysRevLett.121.086602, Nature-Go_OHE, Gambardella-PhysRevLett.131.156703,Manchon-PhysRevB.108.075427, Anderson-PhysRevB.108.245105, Manchon-PhysRevB.106.104414, Kawakami-PhysRevLett.131.156702, Faridi2025, Costa2025, Abrao2025, Veneri2025, Canonico-Roche-PhysRevB.110.L140201}, orbital Edelstein effect \cite{Us-PhysRevB.107.115402, Us-PhysRevB.104.165403, Johansson2024, Orbital-Edelstein-PhysRevResearch.3.013275, Murakami-doi:10.1021/acs.nanolett.7b04300, Hayami2018, Hayami-OME-JPCM, Chirolli-PhysRevLett.128.217703,Fert-doi:10.1021/acs.nanolett.4c01607}, and orbital torque \cite{Go-PhysRevResearch.2.013177, Go-10.1038/s42005-023-01139-7, Canonico2025, Azevedo-PhysRevApplied.19.014069, Hayashi2023, Ando2023}, which together have given rise to the emerging field of orbitronics. Beyond the practical appeal of manipulating electronic OAM for storing and processing information, orbitronics has enabled fundamental advances in understanding the topological aspects of solids \cite{Cysne-PhysRevLett.126.056601, Cysne2025, Orb-Chern-PhysRevB.109.155407, Us-PhysRevLett.130.116204, Yao2025, Medina2025}.

The unbounded nature of the position operator in a periodic solid \cite{Resta-PhysRevLett.80.1800, Blount1962} has posed difficulties in understanding electronic OAM. For this reason, the vast majority of studies in orbitronics have made use of the intra-atomic approximation \cite{Us-PhysRevB.101.161409, Us-PhysRevB.108.165415, Oppeneer-PhysRevB.106.024410, Oppeneer-PhysRevMaterials.6.095001, Kelly-PhysRevB.109.214427, Zhou-OrbitalPhotocurrent, rang2024orbitalhalleffecttransition, Sayantika-PhysRevB.101.121112}. Within this approximation, the electronic OAM operator of the solid is constructed from the atomic operators of its constituents. This approximation is also known as the atom-centred approximation (ACA) and is particularly well-suited to establishing an intuitive picture of some orbitronic phenomena. For instance, the emerging OAM textures \cite{Beaulieu-PhysRevLett.125.216404, Go-PhysRevLett.121.086602} in the band structure of non-centrosymmetric systems plays a role in orbitronic phenomena, akin to the role of spin texture in spintronic phenomena.

In real systems, closed electron trajectories encompassing many atoms contribute to the OAM of the solid. These contributions are called inter-site contributions. For magnetic systems, inter-site contributions are explicitly taken into account in the modern theory of orbital magnetization \cite{ModernTheory-PhysRevLett.95.137204, ModernTheory-PhysRevLett.95.137205}. This theory has improved the agreement between theoretical predictions and experimental measurements of equilibrium magnetization in certain simple materials \cite{Ceresoli2010}. It has also been shown to have a more significant impact on the orbital magnetization of complex materials \cite{Hanke-PhysRevB.94.121114, Lage2025}.

Bhowal and Vignale \cite{BhowalVignale} took the first step beyond the intra-atomic approximation in the context of the \emph{non-equilibrium} OAM transport. They used the formula for the orbital magnetic moment (OMM) of Bloch states, originally derived by Kohn \cite{Bloch-Orbital-Moment-Khon-1} for isolated bands, to define an orbital current operator \cite{BhowalVignale}. This formula for the OMM was later rederived using the semiclassical wave packet formalism \cite{Bloch-Orbital-Moment-Chang-2}. In general, the OMM has a non-Abelian (matrix) structure, as discussed in Refs. \cite{Chang-Niu-Review_2008, Culcer-Niu-PhysRevB.72.085110} for the case of nearly degenerate bands. In many applications \cite{Wozniak2020, Junior2025}, the diagonal matrix elements (i.e., the expectation values of the OMM operator) should govern the principal physical aspects. For this reason, the matrix structure of the OMM was usually not considered in the literature. With the advent of orbitronics and the study of non-equilibrium orbital currents, the complete matrix structure of the OMM was found to play a crucial role in describing the OHE in centrosymmetric systems \cite{Cysne-Vignale-Tatiana-PhysRevB.105.195421, Cysne2024a}. In principle, the approach proposed by Bhowal and Vignale to describe the OHE accounts for contributions to the OAM arising from the extended nature of electronic wave functions (intersite contributions). This allows the OHE to occur even in solids composed of $s-$ and $p_z$-orbitals in graphene-like systems that lack intra-atomic contributions \cite{BhowalVignale, Cysne-PhysRevResearch.6.023271, Mertig-PhysRevResearch.5.043052}. Nevertheless, caution must be taken when computing non-diagonal matrix elements of the OMM operator. If the usual formula found in the literature is applied to non-degenerate bands, it loses its gauge invariance, as pointed out in Ref. \cite{IvoSousa-10.21468/SciPostPhys.14.5.118}.

A complete formulation of the OMM operator for Bloch electrons is still being developed, along with studies of its implications for the OHE \cite{liu2024quantumcorrectionorbitalhall}. Rigorously, this operator should be applicable to the entire Hilbert space (i.e., not constrained to nearly degenerate coupled bands) while maintaining gauge covariance. It should, in principle, account for both inter-site and intra-site contributions to electronic OAM. Here, we present a consistent derivation of the OMM operator matrix elements for Bloch states and identify a new contribution previously neglected in studies of the OHE. The explicit formula for these matrix elements can be found in Eqs. (\ref{OMMFull}-\ref{gauge2}) in the main text. We study the impact of the new term on the OHE in the bilayer of a 2H transition metal dichalcogenide (TMD) \cite{Cysne-PhysRevLett.126.056601, Cysne-Vignale-Tatiana-PhysRevB.105.195421} and in biased bilayer graphene \cite{Cysne-PhysRevResearch.6.023271}. In both cases, the new contributions cause a significant reduction in the orbital Hall conductivity plateau.

\section{Deriving the formula for the orbital magnetic moment of Bloch states} 

Formally, the $z$-component of the electronic OMM operator is given by $-\frac{e}{4}\left[ \left(\hat{\vec{r}}\times \hat{\vec{v}}\right)-\left(\hat{\vec{v}}\times \hat{\vec{r}} \right)\right]_z$. In this section, we aim to derive an expression for the matrix elements of this operator. Let $\ket{n_{{\bf k}}}$ and $\ket{n'_{{\bf k}}}$ be the periodic parts of the Bloch states of the Hamiltonian $\hat{H}_{\bf k}$ associated with energy bands $\epsilon_{n{\bf k}}$ and $\epsilon_{n'{\bf k}}$. We define the matrix elements:
\begin{eqnarray}
m_{nn'{\bf k}}=-\frac{e}{4}\bra{n}\left[(\hat{\vec{r}}\times \hat{\vec{v}})-(\hat{\vec{v}}\times \hat{\vec{r}}) \right]_z\ket{n'}. \label{mPhys}
\end{eqnarray}
Throughout this paper, we assume a { \it non-degenerate band structure} for the Hamiltonian $\hat{H}_{\bf k}$. The generalization to include degeneracies is left for future work. In this section, to maintain a simple notation, we omit the dependence on ${\bf k}$ in energy and Bloch states, representing them with $\epsilon_{n(n')}$ and $\ket{n(n')}$. 

The intrinsic OMM in the case of isolated (single) Bloch bands was first derived by Kohn \cite{Bloch-Orbital-Moment-Khon-1}. The matrix elements of the OMM operator can be calculated using two distinct approaches. The first consists of computing the matrix elements of the quantum mechanical position and velocity operators in Bloch states \cite{Manchon-PhysRevB.106.104414, Cysne-Vignale-Tatiana-PhysRevB.105.195421, BhowalVignale, Mertig-PhysRevResearch.5.043052}. As we shall discuss, this computation requires caution as important elements may have been overlooked in previous calculations. The second approach is based on the semiclassical formalism, which assumes the construction of a coherent electronic wave packet \cite{Chang-Niu-Review_2008, Culcer-Niu-PhysRevB.72.085110, Bloch-Orbital-Moment-Chang-2}. This work revisits the topic and addresses elements omitted in previous studies. This leads to novel contributions to the OMM that can influence the OHE in certain situations.

\subsection{The generalized semi-classical formula}

The semiclassical approach allows a transparent physical interpretation of the electronic OMM. In this context, the possibility of constructing a semi-classical electronic wave packet is assumed, and the OMM is interpreted as the self-rotation of this coherent state \cite{Bloch-Orbital-Moment-Chang-2, Chang-Niu-Review_2008, Culcer-Niu-PhysRevB.72.085110}. Nevertheless, the Bloch states used to build this wave packet should be restricted to nearly degenerate bands. A previous derivation, based on the evolution of the quantum mechanical position and velocity operators’ matrix elements, yields a similar formula for the OMM but involves the complete Hilbert space \cite{Manchon-PhysRevB.106.104414}. As will be discussed in detail, in addition to losing the interpretation of self-rotation of the wave packet, this generalized formula for the OMM also loses gauge invariance when extended to the complete Hilbert space. In this generalization, the expression for OMM matrix elements is given by:
\begin{eqnarray}
m^{\rm SR}_{nn'{\bf k}}=-\frac{ie}{2\hbar}\bra{\vec{\nabla}_{\bf k}n}\boldsymbol{\times} \left[\hat{H}_{\bf k}-\left(\frac{\epsilon_n+\epsilon_{n'}}{2}\right)\mathbb{1}\right]\ket{\vec{\nabla}_{\bf k}n'}, \nonumber \\
\label{mBloch}
\end{eqnarray}
where $\vec{\nabla}_{\bf k}=\vec{x}(\partial/\partial k_{x})+\vec{y}(\partial/\partial k_{y})$. Note that we keep the superscripts S.R., although Eq.(\ref{mBloch}) can no longer be interpreted as a self-rotation.

To evaluate Eq. (\ref{mBloch}), we make use of the following identity for the derivative of the Bloch state, valid for non-degenerate spectra:
\begin{eqnarray}
\ket{\vec{\nabla}_{\bf k}n}= \sum_{m\neq n}\frac{\hbar \bra{m}\hat{\vec{v}}_{\bf k}\ket{n}}{\epsilon_n-\epsilon_m}\ket{m}-i \vec{\mathcal{A}}_{n{\bf k}}\ket{n}, \label{I1Ket}
\end{eqnarray}
where the Berry connection of a Bloch band is expressed as $\vec{\mathcal{A}}_{n{\bf k}}=i\braket{n|\vec{\nabla}_{\bf k}n}=\mathcal{A}^x_{n{\bf k}}\vec{x}+\mathcal{A}^y_{n{\bf k}}\vec{y}$. The demonstration of Eq. (\ref{I1Ket}) is presented in Appendix \ref{AppA} (see also Chapter 2 of Ref. \cite{Vanderbilt2018}). The Berry connection term on the right-hand side of Eq. (\ref{I1Ket}) plays an important role in evaluating the OMM matrix elements between states with distinct energies, as it restores gauge covariance. In typical studies of orbital magnetism, where only the diagonal matrix elements of the OMM operator matter, the Berry connection term in identity of Eq. (\ref{I1Ket}) does not affect the result. However, in the case of non-equilibrium orbital magnetic moment flow via the OHE, non-diagonal elements play a crucial role \cite{Cysne-Vignale-Tatiana-PhysRevB.105.195421, Cysne2024a}, and it is important to include the Berry connection term. By inserting Eq. (\ref{I1Ket}) into Eq. (\ref{mBloch}), we obtain
\begin{widetext}
\begin{eqnarray}
m^{\rm SR}_{nn'{\bf k}}=-\frac{ie\hbar}{2}&&\Bigg\{\sum_{\substack{m\neq n \\ \tilde{m}\neq n'}} \frac{v^x_{nm} v^y_{\tilde{m}n'} \left(\epsilon_m-\bar{\epsilon}_{nn'}\right)\delta_{m\tilde{m}}}{\Delta\epsilon_{nm}\Delta\epsilon_{n'\tilde{m}}}+i\frac{\mathcal{A}^x_{n{\bf k}}}{\hbar} \left(\epsilon_n-\bar{\epsilon}_{nn'}\right) \sum_{\tilde{m}\neq n'} \frac{v^y_{\tilde{m}n'}\delta_{n\tilde{m}}}{\Delta\epsilon_{n'\tilde{m}}}-i\frac{\mathcal{A}^y_{n'{\bf k}}}{\hbar} \left(\epsilon_{n'}-\bar{\epsilon}_{nn'}\right) \sum_{m\neq n} \frac{v^x_{nm}\delta_{mn'}}{\Delta\epsilon_{nm}}  \nonumber \\
&&-\left(x\leftrightarrow y\right)\Bigg\},
\end{eqnarray}
\end{widetext}
where we used the notation $v^{\mu}_{\gamma \beta}=\bra{\gamma}\hat{v}^{\mu}_{\bf k}\ket{\beta}$, $\Delta\epsilon_{\gamma \beta}=\left(\epsilon_{\gamma}-\epsilon_{\beta}\right)$, and $\bar{\epsilon}_{nn'}=\left(\epsilon_n+\epsilon_{n'}\right)/2$. In the previous equation, a term $\mathcal{A}^x_{n{\bf k}}\mathcal{A}^y_{n'{\bf k}} \delta_{nn'} \left(\epsilon_n-\bar{\epsilon}_{nn'} \right)/\hbar^2$ is absent because $\delta_{nn'} \left(\epsilon_n-\bar{\epsilon}_{nn'}\right)=0$. For the same reason, the second and third terms are non-zero only when $n\neq n'$. Then we can use $\sum_{\tilde{m}\neq n'}\frac{v^y_{\tilde{m}n'}\delta_{n\tilde{m}}}{\Delta\epsilon_{n'\tilde{m}}}=\frac{v^y_{nn'}}{\Delta\epsilon_{n'n}}$. After performing some algebra, we obtain the following result:
\begin{eqnarray}
m^{\rm SR}_{nn'{\bf k}}=F_{nn'{\bf k}}+g_{nn'{\bf k}}, \label{mBlochExpr}
\end{eqnarray}
where
\begin{eqnarray}
F_{nn'{\bf k}}=\frac{ie\hbar}{2}\sum_{\substack{m\neq n \\ \tilde{m}\neq n'}}\Bigg\{\frac{v^x_{nm} v^y_{\tilde{m}n'} \left(\bar{\epsilon}_{nn'}-\epsilon_m\right)\delta_{m\tilde{m}}}{\left(\epsilon_n-\epsilon_m\right)\left(\epsilon_{n'}-\epsilon_{\tilde{m}}\right)}-(x\leftrightarrow y) \Bigg\}, \nonumber \\ \label{Fnnp}
\end{eqnarray}
and
\begin{eqnarray}
&&g_{nn'{\bf k}}=-\frac{e}{4}\Big[\left(\mathcal{A}^x_{n{\bf k}}v^y_{nn'}-\mathcal{A}^y_{n'{\bf k}}v^x_{nn'}\right)-(x\leftrightarrow y)\Big](1-\delta_{nn'}). \nonumber \\ 
&&\label{gnnp}
\end{eqnarray}
Eqs. (\ref{mBlochExpr}-\ref{gnnp}) correspond exactly to Eq. (61) of Ref. \cite{IvoSousa-10.21468/SciPostPhys.14.5.118} obtained under the premise that the OMM matrix element must be gauge covariant. The term $g_{nn'{\bf k}}$ originates from the Berry connection term that appears on the right-hand side of Eq. (\ref{I1Ket}).

\subsection{OMM from matrix elements of ${\bf \hat{r}}$ and ${\bf \hat{v}}$ operators \label{SecmPhys}}

We now evaluate Eq. (\ref{mPhys}) by explicitly computing the matrix elements of the quantum-mechanical position and velocity operators in Bloch states. This calculation must be performed consistently with the identity introduced in the previous subsection. The matrix element of the quantum-mechanical position operator between two non-degenerate Bloch states is written as follows \footnote{Eq. (\ref{rnnl}) should not be over-interpreted as the crystal-momentum representation of the position operator, whose correct formulation has been sought for a long time \cite{Blount1962, Resta1998} and still generates debate \cite{Si2025}. Instead, it should be seen as an expression consistent with Eq.(\ref{I1Ket}), as proved in the Appendix \ref{AppA}, with the latter being compatible with the covariant derivative of eigenstates.}: 
\begin{eqnarray}
\bra{n}\hat{\vec{r}}\ket{n'}&=&\bra{n}\left(i\hat{\vec{d}}-\mathbb{1}\vec{\mathcal{A}}_{n'{\bf k}}\right)\ket{n'} \nonumber \\
&=& i\braket{n|\vec{\nabla}_{\bf k}n'}-\delta_{nn'}\vec{\mathcal{A}}_{n'{\bf k}}.
\label{rnnl}
\end{eqnarray}
An analogous expression was introduced in Ref. \cite{Blount1962} and has been widely used in the literature \cite{BhowalVignale, SunVignale-2024-OrbitalMagnetoresistance}. The consistency between Eq. (\ref{rnnl}) and the identity of Eq. (\ref{I1Ket}) is demonstrated in Appendix \ref{AppA}. Once again, the Berry connection term plays an important role in the OMM matrix elements involving non-degenerate states.

\begin{widetext}
To cast the OMM matrix element in a convenient form, we write $(\hat{\vec{r}}\times\hat{\vec{v}})_z-(\hat{\vec{v}}\times\hat{\vec{r}})_z=\hat{x}\hat{v}^y+\hat{v}^y\hat{x}-\left(x \leftrightarrow y \right)$.
Then, we insert the resolution of identity into Eq. (\ref{mPhys}) and use Eq. (\ref{rnnl}) to represent the matrix elements of the position operators:
\begin{eqnarray}
m_{nn'{\bf k}}&=&-\frac{e}{4}\sum_m\Big[\bra{n}\hat{x}\ket{m}\bra{m}\hat{v}^y\ket{n'}+\bra{n}\hat{v}^y\ket{m}\bra{m}\hat{x}\ket{n'}-(x \leftrightarrow y)\Big] \nonumber \\
&=&-\frac{e}{4}\sum_m\Big[-i\braket{\partial_x n|m}v^y_{mn'}+iv^y_{nm}\braket{m|\partial_xn'}-(x\leftrightarrow y)\Big]+\frac{e}{4}\Big[\mathcal{A}^x_{n{\bf k}}v^y_{nn'}+v^y_{nn'}\mathcal{A}^x_{n'{\bf k}}-(x\leftrightarrow y)\Big] \nonumber \\
&=&F_{nn'{\bf k}}+\tilde{g}_{nn'{\bf k}}, \label{mPhysIandII}
\end{eqnarray}
where $F_{nn'{\bf k}}$ is given by Eq.(\ref{Fnnp}) and,
\begin{eqnarray}
\tilde{g}_{nn'{\bf k}}=\frac{e}{4}\left[v^y_{n'n'}i\braket{\partial_xn|n'}+v^x_{nn}i\braket{n|\partial_yn'}\right]\left(1-\delta_{nn'}\right)-(x\leftrightarrow y).
\label{gtilde}
\end{eqnarray}
The third line in Eq.(\ref{mPhysIandII}) follows from straightforward steps outlined in Appendix \ref{APPmathDetails}. Note that $\tilde{g}_{nn'{\bf k}}$ originates from the Berry connection term on the right-hand side of Eq.(\ref{rnnl}).
\end{widetext}

\subsection{Relation Between $m_{nn'{\bf k}}$ and $m^{\rm SR}_{nn'{\bf k}}$}
Now we can establish the relation between the electronic OMM matrix element, defined via the position and velocity operators, and the expression analogous to the semiclassical formula commonly used in the literature. By isolating the function $F_{nn'{\bf k}}$ in Eqs.(\ref{mBlochExpr}) and (\ref{mPhysIandII}), one finds the relation: 
\begin{eqnarray}
m_{nn'{\bf k}}=m^{\rm SR}_{nn'{\bf k}}+f_{nn'{\bf k}}, \label{FinalRelation1}
\end{eqnarray}
where $f_{nn'{\bf k}}=\tilde{g}_{nn'{\bf k}}-g_{nn'{\bf k}}$. Using expressions of Eqs. (\ref{gnnp}) and (\ref{gtilde}), and $\braket{\partial_{x(y)}n|n'}=-\braket{n|\partial_{x(y)}n'}$ (valid for $n\neq n'$) is obtained, after some straightforward manipulations:
\begin{eqnarray}
f_{nn'{\bf k}}&=&\frac{e}{4}\left[\left(\mathcal{A}^x_{n{\bf k}}+\mathcal{A}^x_{n'{\bf k}}\right)v^y_{nn'}+\left( v^x_{nn}+ v^x_{n'n'}\right)i\braket{n|\partial_y n'}\right] \nonumber \\
&&.(1-\delta_{nn'})-(x\leftrightarrow y), \label{fFinal}
\end{eqnarray}
which is the final result of this section. In the next section, we discuss separately the mathematical structure and physical meaning of each term in these equations.

\section{Complete formula for the electronic OMM}
In the previous section, we have shown that the matrix elements of the OMM operator of Bloch electrons between non-degenerate states, $m_{nn'{\bf k}}=-\frac{e}{4}\bra{n}\left[(\hat{\vec{r}}\times \hat{\vec{v}})-(\hat{\vec{v}}\times \hat{\vec{r}}) \right]_z\ket{n'}$, can be written as:
\begin{eqnarray}
m_{nn'{\bf k}}=m^{\rm SR}_{nn'{\bf k}}+g^{\rm I}_{nn'{\bf k}}+g^{\rm II}_{nn'{\bf k}}, \label{OMMFull}
\end{eqnarray}
where, $m^{\rm SR}_{nn'{\bf k}}$ is analogous to the term obtained in the semiclassical wave packet approach:
\begin{eqnarray}
m^{\rm SR}_{nn'{\bf k}}=-\frac{ie}{2\hbar}\bra{\vec{\nabla}_{\bf k}n}\boldsymbol{\times} \left[\hat{H}_{\bf k}-\left(\frac{\epsilon_n+\epsilon_{n'}}{2}\right)\mathbb{1}\right]\ket{\vec{\nabla}_{\bf k}n'}, \nonumber \\
\label{mSR}
\end{eqnarray}
where $\vec{\nabla}_{\bf k}=\vec{x}(\partial/\partial k_{x})+\vec{y}(\partial/\partial k_{y})$ are ordinary derivatives and $\boldsymbol{\times}$ is the cross product. The semiclassical approach consists of constructing a coherent electronic wave packet. The term analogous to Eq. (\ref{mSR}) is often interpreted as the magnetic moment arising from the self-rotation of this wave packet \cite{Chang-Niu-Review_2008}.
Here, we are dealing with the full quantum mechanical treatment. In contrast to the semi-classical approach \cite{Culcer-Niu-PhysRevB.72.085110}, the non-zero matrix elements of the equation are not restricted to the quasi-degenerate Hilbert subspace. Consequently, as pointed out in Ref. \cite{IvoSousa-10.21468/SciPostPhys.14.5.118}, Eq.(\ref{mSR}) alone is not gauge-covariant. The term $g^{\rm I}_{nn'{\bf k}}$ is given by 
\begin{eqnarray}
g^{\rm I}_{nn'{\bf k}}=\hspace{3mm}\frac{e}{4}\left[\left(\vec{\mathcal{A}}_{n{\bf k}}+\vec{\mathcal{A}}_{n'{\bf k}}\right)\boldsymbol{\times} \vec{v}_{nn'}\right]_z (1-\delta_{nn'}).
\label{gauge1}
\end{eqnarray}
As discussed in Ref. \cite{IvoSousa-10.21468/SciPostPhys.14.5.118}, the quantity $m^{\rm SR}_{nn'{\bf k}}+g^{\rm I}_{nn'{\bf k}}$ is gauge-covariant and takes a form similar to Eq. (\ref{mSR}), with ordinary derivatives of eigenstates $\ket{\vec{\nabla}_{{\bf k}}n}$ replaced by covariant derivatives $\ket{{\bf D}_{\bf k}n}=\ket{\vec{\nabla}_{{\bf k}}n}+i\vec{\mathcal{A}}_{n{\bf k}}\ket{n}$. Here, we obtain a second contribution:
\begin{eqnarray}
g^{\rm II}_{nn'{\bf k}}=\frac{e}{4}\Big[\big(\vec{v}_{nn}+\vec{v}_{n'n'}\big)\boldsymbol{\times} i\braket{n|\vec{\nabla} n'}\Big]_z (1-\delta_{nn'}). \nonumber \\
\label{gauge2}
\end{eqnarray}
The same expressions of Eqs. (\ref{OMMFull}-\ref{gauge2}) were recently derived through a multipole expansion of the optical matrix elements of crystals, a derivation that differs completely from the one presented here (see Appendix A of Ref. \cite{Urru2025}). To our knowledge, the new term in Eq. (\ref{gauge2}) for the OMM has not been considered in previous works on the OHE and constitutes one of the main results of this paper. It is worth mentioning that the quantity $m^{\rm SR}_{nn'{\bf k}}+g^{\rm I}_{nn'{\bf k}}+g^{\rm II}_{nn'{\bf k}}$ remains gauge-covariant and therefore has physical significance, being rigorously interpreted as the OMM of Bloch electrons. It is worth noting that in the derivation of Eqs. (\ref{OMMFull}-\ref{gauge2}), we assumed a Hamiltonian with a non-degenerate electronic spectrum. 

By introducing multiplicative constants as done in Ref. \cite{BhowalVignale} and making use of Eq. (\ref{I1Ket}) in the equations above, one can express the OAM matrix elements of Bloch electrons as a sum over states \cite{Urru2025}:
\begin{eqnarray}
L^z_{nn'{\bf k}}=-\frac{ie\hbar^2}{4g_{\rm L}\mu_{\rm B}}&&\Bigg\{\sum_{m\neq n'}\frac{\vec{v}_{nm}\boldsymbol{\times}\vec{v}_{mn'}}{\epsilon_{n'}-\epsilon_{m}} + \sum_{m\neq n}\frac{\hat{\vec{v}}_{nm}\boldsymbol{\times}\hat{\vec{v}}_{mn'}}{\epsilon_n-\epsilon_m}\Bigg\}, \nonumber \\
\label{SumStates}
\end{eqnarray}
where $g_{\rm L}=1$, $\mu_{\rm B}=(e\hbar)/(2m_e)$, $\vec{v}_{\gamma\beta}=\bra{\gamma}\hat{\vec{v}}_{\bf k}\ket{\beta}$ and $\boldsymbol{\times}$ is the cross product. This formula is better suited for numerical computation of the OAM matrix elements \cite{FariaJunior2025}.

In the following sections, we construct the full OMM for Bloch electrons and examine the implications of the new terms for the OHE, using the low-energy theory of bilayer 2H-TMD and biased bilayer graphene as illustrative examples.

\section{Impact of the $\boldsymbol{g^{\rm II}}$ Term on the OHE in bilayer 2H-TMD \label{OHE2HTMD}}
\subsection{Low energy theory for bilayer of 2H-TMD}

In the low-energy regime, the Hamiltonian of a bilayer TMD in the 2H structural phase can be written as \cite{Low-Energy-bilayer, Bukard-PhysRevB.98.035408}
\begin{eqnarray}
H^{\rm TMD}_{{\bf q}_{\tau}}=\begin{bmatrix}
\Delta & \gamma_+  & 0 & 0\\
  \gamma_- & 0 & 0 & t_{\perp}\\
0 & 0 & \Delta & \gamma_-\\
0 & t_{\perp} & \gamma_+ & 0
\end{bmatrix},
\label{Heff}
\end{eqnarray}
where  $\gamma_\pm = at(\tau q_x \pm i q_y)=\tau at q e^{\pm i \tau \theta} $ and $q=\sqrt{q^2_x+q^2_y}$. $\tau=\pm 1$ is the valley quantum number associated with valleys ${\bf K}=(4\pi/3a){\bf x}$ and ${\bf K}'=-{\bf K}$, respectively. The tight-biding basis of the Hamiltonian of Eq. (\ref{Heff}) is $\beta_{tb}=\{ \big|d^1_{z^2}\big>,\big( \big|d^1_{x^2-y^2}\big>-i\tau \big|d^1_{xy}\big> \big)/\sqrt{2}, \big|d^2_{z^2}\big>, \big( \big|d^2_{x^2-y^2}\big>+i\tau \big|d^2_{xy}\big> \big)/\sqrt{2}\}$, where the superscripts 1 and 2 specify the two layers of the bilayer, respectively. Here, ${\bf q}_{\tau}={\bf k}-\tau {\bf K}$ where ${\bf q}_{\tau}$ represents the wavevector relative to valleys. This model can be easily applied to describe bilayers of compounds of the TMD family in the trigonal prismatic phase (H), with 2H stacking. In this paper, we will consider the case of bilayers of 2H-MoS$_2$. The parameters of effective Hamiltonian can be obtained by adjusting the results given by density functional theory calculations. For 2H-MoS$_2$ bilayers, we obtain the band-gap $\Delta=1.766 \text{eV}$, the lattice constant $a=3.160 \angstrom$, the intralayer nearest-neighbor hopping $t=1.137$ eV, and the interlayer hopping $t_{\perp}=0.043$ eV \cite{Low-Energy-bilayer, Bukard-PhysRevB.98.035408}. Previous studies have shown that the spin-orbit interaction plays a minor role in the transport of OAM in TMDs \cite{Us-PhysRevB.101.161409, Bhowal2020}. Since we focus on the orbital degrees of freedom, we neglect the spin-orbit coupling and introduce a spin-degeneracy factor $g_s=2$. Here, we consider the case of the unbiased bilayer, with the Hamiltonian of Eq. (\ref{Heff}) respecting spatial-inversion symmetry ($\mathcal{P}\rightarrow\checkmark$). Clearly, Eq. (\ref{Heff}) also respects time-reversal symmetry ($\mathcal{T}\rightarrow\checkmark$). In Appendix \ref{AppB}, we present the energies and wave functions of the bilayer system, corrected to first order in interlayer hopping.

\subsection{Orbital magnetic moment operator}

Here, we construct the OMM operator for bilayer 2H-TMD, calculating explicitly each term in Eq. (\ref{OMMFull}) using the perturbed vector states and energies from Eqs. (\ref{CorrectedEnergies}) and (\ref{CorrectedStates}), respectively. Note that interlayer coupling lifts the degeneracy of the energy spectrum in the 2H-TMD bilayer. Consequently, Eq. (\ref{OMMFull}) can be applied. Using the basis $\beta_{\Phi}=\{\Phi_{-v}; \Phi_{+v}; \Phi_{-c}; \Phi_{+c}\}$ [Eq.(\ref{CorrectedStates})] to express the matrices we obtain the first contribution [Eq.(\ref{mSR})] given:
\begin{eqnarray}
\hat{\mathbb{m}}_{{\bf q}_{\tau}}^{{\rm SR}}= \tau\begin{bmatrix}
0 & m^0_q & 0 & -w^{+}_q\\
m^0_q  & 0 & -w^{-}_q & 0 \\
0 & -w^{-}_q & 0 & m^0_q \\
-w^{+}_q & 0 & m^0_q  & 0 
\end{bmatrix}. \nonumber \\ \label{MMSRtmd}
\end{eqnarray}
where, $m^0_q=\left(\frac{e}{\hbar}\right) \frac{a^2 t^2\Delta}{4a^2t^2q^2+\Delta^2}$ and $w^{\pm}_q=\left(\frac{e}{\hbar}\right)\frac{\Delta ^5+16 \Delta  q^4 t^4a^4+8 \Delta ^3 q^2 t^2a^2\pm t_{\perp}\Delta^3 \sqrt{\Delta ^2+4 q^2 t^2a^2}}{4 q^2 \left(\Delta ^2+4 q^2 t^2a^2\right)^2 \sqrt{\frac{4q^2a^2t^2+\Delta^2}{q^2a^2t^2}}}$. We briefly mention that in a previous work by some of us \cite{Cysne-Vignale-Tatiana-PhysRevB.105.195421}, the construction of the OMM operator was based on a semi-classical approach. In that case, the terms $w^{\pm}_q$, which couple the conduction and valence band states, were not considered \cite{Culcer-Niu-PhysRevB.72.085110, Chang-Niu-Review_2008}. Nevertheless, their influence on the final expression for orbital Hall conductivity disappears after the azimuthal integration of orbital Berry curvatures (see discussion in Appendix A of Ref. \cite{Cysne-Vignale-Tatiana-PhysRevB.105.195421}). The term of Eq.(\ref{gauge1}) for states of Eq. (\ref{CorrectedStates}) is identically zero, 
\begin{eqnarray}
\hat{\mathbb{g}}^{\rm I}_{{\bf q}_{\tau}}=\mathbb{0}_{4\times4}. 
\end{eqnarray}
Finally, the new term of Eq.(\ref{gauge2}) results in a non-zero contribution that  affects the final expression for the orbital Hall conductivity of bilayer TMD:
\begin{eqnarray}
\hat{\mathbb{g}}^{\rm II}_{{\bf q}_{\tau}}= \tau\begin{bmatrix}
0 & m^{\rm II+}_{q} & 0 & 0\\
m^{\rm II+}_{q}  & 0 & 0 & 0 \\
0 & 0 & 0 & m^{\rm II-}_{q} \\
0 & 0 &m^{\rm II-}_{q} & 0 
\end{bmatrix}, \nonumber \\ \label{MMg2tmd}
\end{eqnarray}
where $m^{\rm II\pm}_{q}=\left(\frac{e}{\hbar}\right)\frac{a^2t^2 \left(-\Delta\pm\sqrt{\Delta ^2+4 q^2 a^2t^2} \right)}{2 \left(\Delta ^2+4 q^2 a^2t^2\right)}$. In the basis $\beta_{\Phi}$, the full OMM matrix is the sum of the previous three contributions $\hat{\mathbb{m}}_{{\bf q}_{\tau}}=\hat{\mathbb{m}}_{{\bf q}_{\tau}}^{\rm SR}+\hat{\mathbb{g}}_{{\bf q}_{\tau}}^{\rm I}+\hat{\mathbb{g}}_{{\bf q}_{\tau}}^{\rm II}$. 

To use the OMM operator in the Kubo formula, we first transform it into the tight-binding basis of the Hamiltonian of Eq. (\ref{Heff}) by performing a unitary transformation defined by $U_{{\bf q}_{\tau}}=\{ \beta_{\Phi} \rightarrow \beta_{\rm t.b.}\}$, obtaining: $\hat{\mathbb{m}}^{\rm tb}_{{\bf q}_{\tau}}=U_{{\bf q}_{\tau}} \hat{\mathbb{m}}_{{\bf q}_{\tau}}U^{\dagger}_{{\bf q}_{\tau}}$. Then, we follow Ref. \cite{BhowalVignale} and define the OAM operator for Bloch electrons by introducing a multiplicative constant to express our results in convenient units:
\begin{eqnarray}
\hat{\mathbb{L}}^z_{{\bf q}_{\tau}}=-\frac{\hbar}{g_{\rm L}\mu_{\rm B}} \hat{\mathbb{m}}^{\rm tb}_{{\bf q}_{\tau}}. \label{Lz}
\end{eqnarray}
where $g_{\rm L}=1$ and $\mu_{\rm B}=(e\hbar)/(2m_e)$ is the Bohr magneton and, $m_e$ is the electron rest mass. It should be noted that the above transformation of units is merely a matter of convenience, allowing us to define the orbital Hall conductivity in the same units as the spin Hall conductivity. As discussed in Ref. \cite{Cysne-Vignale-Tatiana-PhysRevB.105.195421}, we could, in principle, describe the same physics in terms of the OMM current.

\subsection{Orbital Hall conductivity}

\begin{widetext}
To study the OHE, we use the formalism of linear-response theory where the orbital Hall current that flows in the $y$-direction with OAM polarized in the $z$-direction (out-of-plane), generated by a longitudinal ($x$-direction) electric field is proportional to the orbital Hall conductivity, $\mathcal{J}^{L_z}_y=\sigma^{L_z}_{\rm OH}\mathcal{E}_x$, where,
\begin{eqnarray}
\sigma^{L_z}_{\rm OH}=e\sum_n\sum_{\tau=\pm 1}\int \frac{d^2{\bf q}}{(2\pi)^2} f_{n{\bf q}_{\tau}} \Omega^{L_z}_{n{\bf q}_{\tau}}. \label{SigmaLR}
\end{eqnarray}
$f_{n{\bf q}_{\tau}}=\Theta (E_{\rm F}-\tilde{E}_{n}(q))$ is the Fermi-Dirac distribution at zero temperature and Fermi energy $E_{\rm F}$. The orbital-weighted Berry curvature is given by
\begin{eqnarray}
\Omega^{L_z}_{n{\bf q}_{\tau}}=2\hbar\sum_{m\neq n} \text{Im} \left[ \frac{ \bra{\Phi_{n}({\bf q}_{\tau})}\hat{v}^{x}_{{\bf q}_{\tau}}\ket{\Phi_{m}({\bf q}_{\tau})} \bra{\Phi_{m}({\bf q}_{\tau})}\hat{J}^{y,L_z}_{{\bf q}_{\tau}}\ket{\Phi_{n}({\bf q}_{\tau})}}{\left(\tilde{E}_{n}(q)-\tilde{E}_{m}(q)\right)^2} \right] . \nonumber \\
\label{OBC}
\end{eqnarray}
In the above equations, $\tilde{E}_{n(m)}(q)$ are the energies [Eq. (\ref{CorrectedEnergies})] of the states $\ket{\Phi_{n(m)}({\bf q}_{\tau})}$ [Eq. (\ref{CorrectedStates})] corrected by the interlayer hopping within first-order perturbation theory (see details in Appendix \ref{AppB}). The indices $n$ and $m$ contain references to the energy bands (conduction $c$ or valence $v$) and to the sign in the linear combinations (bonding $+$ or antibonding $-$) in the corrected states and energies. The velocity operator in the $x$($y$)-direction for electrons in bilayer TMD is defined by $\hat{v}^{x(y)}_{{\bf q}_{\tau}}=\hbar^{-1}\partial \hat{H}^{\rm TMD}_{{\bf q}_{\tau}}/\partial q_{x(y)}$. We define the operator of OAM current flowing in the $y$-direction as $\hat{J}^{y,L_z}_{{\bf q}_{\tau}}=\frac{1}{2}\left[\hat{\mathbb{L}}_{{\bf q}_{\tau}}^z\hat{v}^y_{{\bf q}_{\tau}}+\hat{v}^y_{{\bf q}_{\tau}} \hat{\mathbb{L}}_{{\bf q}_{\tau}}^z\right]$ \cite{BhowalVignale}. 
\end{widetext} 

\begin{figure}[h!]
	\centering
	 \includegraphics[width=1\linewidth,clip]{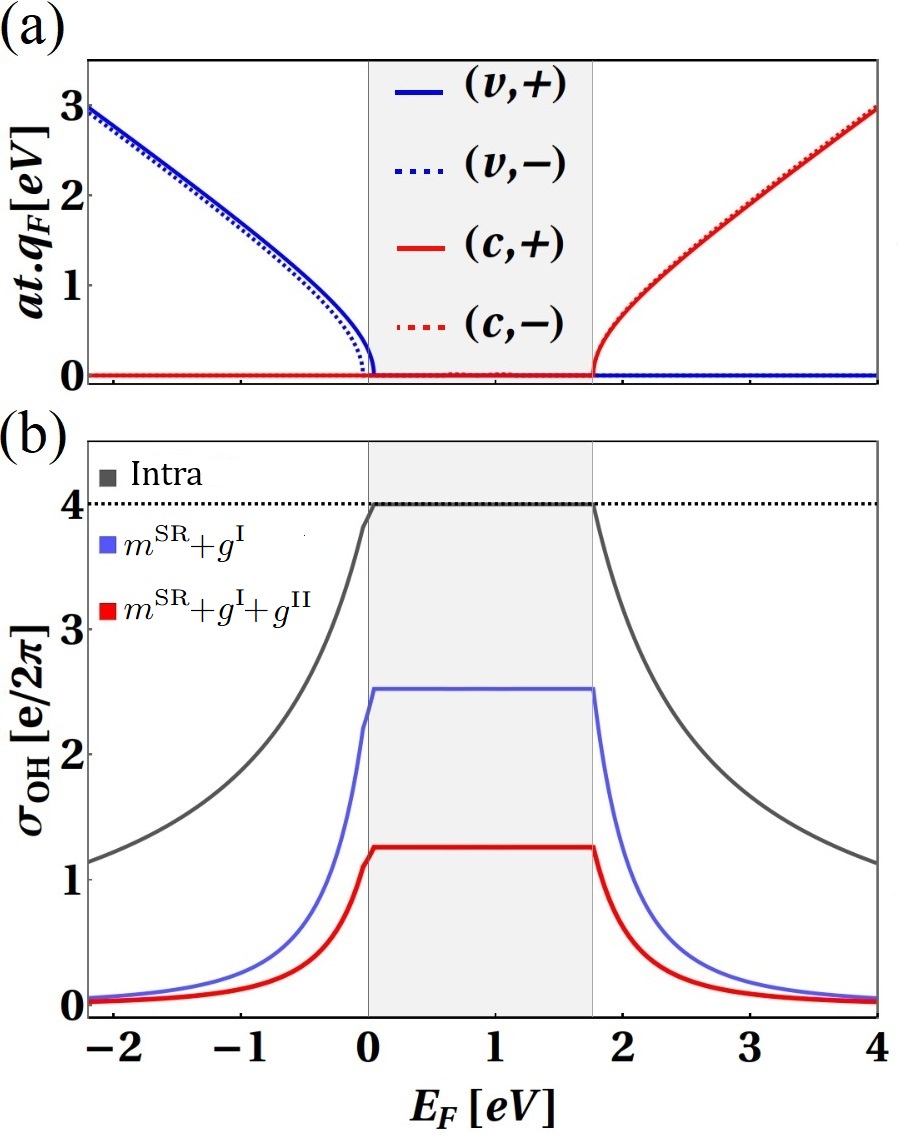}
	\caption{(a) The Fermi momenta of the bilayer 2H-MoS$_2$ bands as a function of Fermi energy. (b) Orbital Hall conductivity of a bilayer 2H-MoS$_2$ as a function of Fermi energy, with the OAM operator described by different approaches: The intra-atomic approximation is shown by the gray curve (see Ref. \cite{Cysne-PhysRevLett.126.056601}). The Bloch state OMM approach, considering only the $m^{\rm SR}_{nn'{\bf k}}+g^{\rm I}_{nn'{\bf k}}$ [Eqs.(\ref{mSR}) and (\ref{gauge1})] (see Ref. \cite{Cysne-Vignale-Tatiana-PhysRevB.105.195421}), is shown by the blue curve. The red curve shows the result including the contribution from Eq. (\ref{gauge2}) in the OMM operator, $m^{\rm SR}_{nn'{\bf k}}+g^{\rm I}_{nn'{\bf k}}+g^{\rm II}_{nn'{\bf k}}$. The shaded rectangle highlights the energy bandgap in the bilayer 2H-MoS$_2$ band structure.}
	\label{FigOHE-TMD} 
\end{figure}

The analytical evaluation of Eqs. (\ref{SigmaLR}) and (\ref{OBC}) for bilayer 2H-TMD, using the complete OMM operator, are presented in the Appendix \ref{AppB}. The red curve in Fig. \ref{FigOHE-TMD} (b) shows the orbital Hall conductivity as a function of Fermi energy for a bilayer of 2H-MoS$_2$, using the expressions of Eqs. (\ref{OHCfinal}-\ref{sigmaLzvFinal}) with the numerical parameters given below Eq.(\ref{Heff}). The height of the orbital Hall conductivity plateau obtained using the complete formula for the OMM of Bloch states [$m^{\rm SR}_{nn'{\bf k}}+g^{\rm I}_{nn'{\bf k}}+g^{\rm II}_{nn'{\bf k}}$] is given by
\begin{eqnarray}
\bar{\sigma}^{L_z}_{\rm 2l, Full}=\frac{a^2t^2}{3\Delta}\left(\frac{g_se^2}{2\pi \hbar \mu_{\rm B}}\right)=\frac{2}{3}\frac{\mu^*_{\rm B}}{\mu_{\rm B}}\left( \frac{e}{2\pi}\right),
\label{barSigmaQuant}
\end{eqnarray}
where, $\mu^*_{\rm B}\propto \Delta^{-1}$ is the renormalized Bohr magneton, which contains the dependence on the band-structure parameters of the TMD. By using the numerical values of the parameters for the case of MoS$_2$, one obtains $\bar{\sigma}^{L_z}_{\rm 2l, Full.}\approx 1.26 \left(e/2\pi\right)$. This result differs by a factor of 1/2 from that obtained in previous work \cite{Cysne-Vignale-Tatiana-PhysRevB.105.195421} using the semiclassical approach $\bar{\sigma}^{L_z}_{\rm 2l, Full.}=(1/2) \bar{\sigma}^{L_z}_{\rm 2l, SR}$ [blue curve in Fig. \ref{FigOHE-TMD} (b)]. This difference arises from the presence of the term $g^{\rm II}_{nn'{\bf k}}$ in the OMM operator. In Fig. \ref{FigOHE-TMD} (b), we also show the result given by the intra-atomic approximation (gray curve), where the height of the orbital Hall insulating plateau assumes a quantized value for bilayers of 2H-TMDs, due to the relationship between the orbital Hall conductivity and the orbital Chern number $\mathcal{C}^{\rm 2l}_L=2$ in our perturbative calculation \cite{Cysne-PhysRevLett.126.056601, Cysne-Vignale-Tatiana-PhysRevB.105.195421}: $\bar{\sigma}^{L_z}_{\rm Intra}=2\mathcal{C}^{\rm 2l}_L(e/2\pi)=4(e/2\pi)$.

\section{Impact of the $\boldsymbol{g^{\rm II}}$ Term on the OHE in biased bilayer graphene \label{OHE2LGr}}

We now investigate the impact of the correction $g^{\rm II}_{nn'{\bf k}}$ on the OMM of Bloch states, considering the case of a bilayer graphene (BLG). In the low-energy limit, we use the tight-binding (TB) basis of the Hamiltonian of BLG with Bernal stacking in the valley ${\bf K}=\hat{x}(4\pi)/(3\sqrt{3}a)$ as $\beta^{ \rm BL}_{tb}({\bf K})=\{ A_u, B_b, A_b, B_u\}$, and in the valley ${\bf K}'=-{\bf K}$ as $\beta^{\rm BL}_{tb} ({\bf K}')=\{ B_b, A_u, B_u, A_b\}$. $A$ and $B$ refer to the sublattices of the honeycomb arrangement, and subscripts $u$ and $b$ refer to the top and bottom layers of the BLG system. In this basis, one writes the low-energy Hamiltonian as \cite{McCann2006}
\begin{eqnarray}
H^{\rm BLG}_{{\bf q}_{\tau}}=\begin{bmatrix}
   -\frac{\tau \Delta}{2} & 0 & 0 & \tilde{\gamma}_{-} \\
   0 & \frac{\tau \Delta}{2} & \tilde{\gamma}_{+} & 0 \\
   0 & \tilde{\gamma}_{-} & \frac{\tau \Delta}{2} & t_{\perp} \\
   \tilde{\gamma}_{+} & 0 & t_{\perp} & -\frac{\tau \Delta}{2}
\end{bmatrix},
\label{HBGB}
\end{eqnarray}
where $\tilde{\gamma}_{\pm}=\hbar \tilde{v}\tau(q_x\pm i q_y)$ with ${\bf q}_{\tau}={\bf k}-\tau {\bf K}$ and $\tilde{v}=3a\tilde{t}/(2\hbar)$ with $a=1.42 \angstrom$ and the renormalized nearest-neighbor hopping amplitude $\tilde{t}= 3.16 \text{eV}$. For the isolated BLG, the interlayer hopping is $t_{\perp}=0.38 \text{eV}$. $\tau$ is the quantum number associated with the valley degree of freedom and assumes values $\tau=\pm 1$ for valleys ${\bf K}$ and ${\bf K}'$, respectively. The term $\Delta$ produces an asymmetry in the on-site energy of each layer, modeling the effect of a perpendicularly applied electric field or the interaction of the BLG with a substrate. The Hamiltonian in Eq. (\ref{HBGB}) respects time-reversal symmetry ($\mathcal{T}\rightarrow\checkmark$) but, in contrast to the model studied in the previous section, $H^{\rm BLG}_{{\bf q}_{\tau}}$ breaks spatial inversion symmetry ($\mathcal{P}\rightarrow\times$) when $\Delta\neq 0$.

\begin{figure}
	\centering
	 \includegraphics[width=1\linewidth,clip]{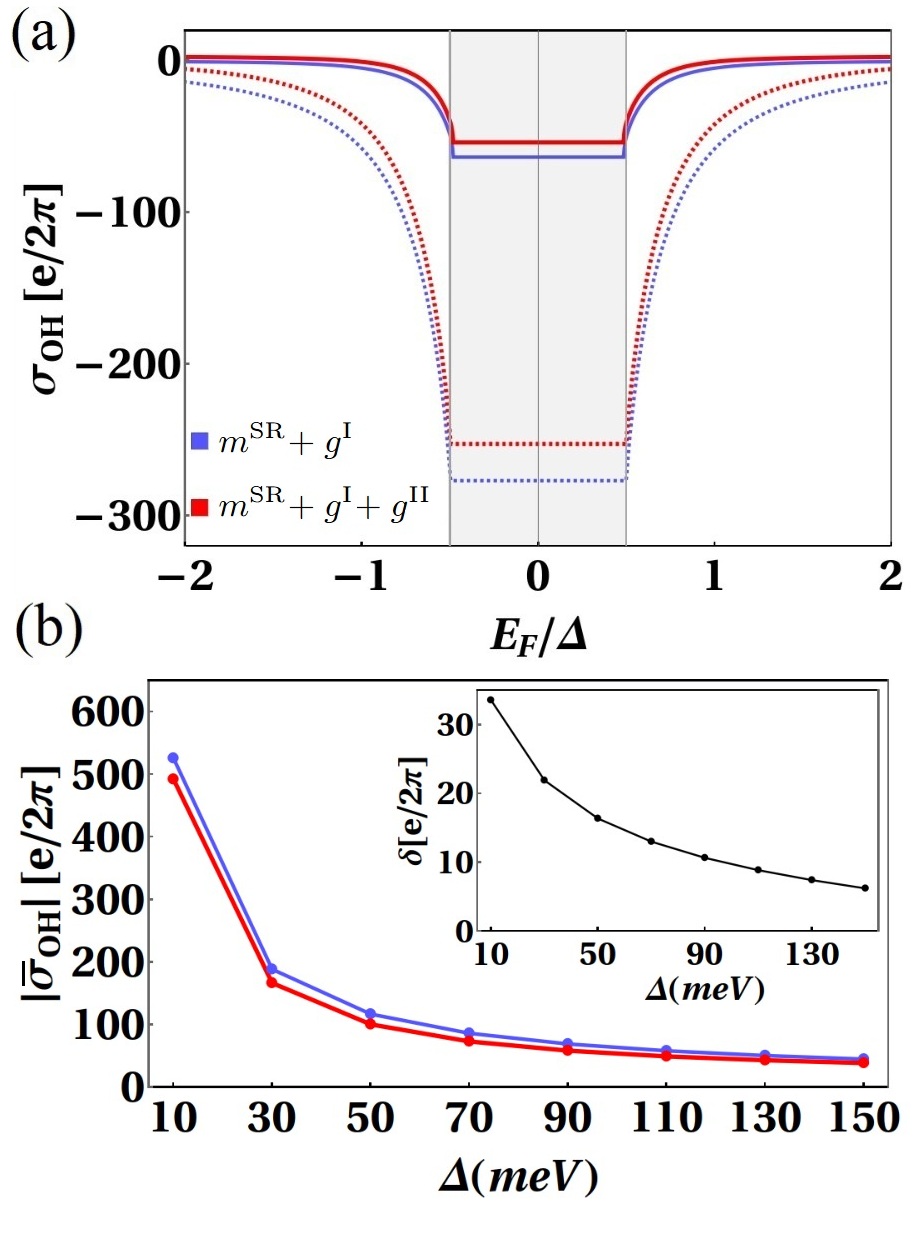}
	\caption{(a) Orbital Hall conductivity of bilayer graphene subjected to an external perpendicular electric field [Eq.(\ref{HBGB})]. Here, we consider two values for the external electric field parameter: $\Delta=100 \text{meV}$ (solid lines) and $20 \text{meV}$ (dashed lines). The blue curve represents the results when only $m^{\rm SR}_{nn'{\bf k}}+g^{\rm I}_{nn'{\bf k}}$ are included in the description of the orbital magnetic moment. Red curves represent the results when the complete description $m^{\rm SR}_{nn'{\bf k}}+g^{\rm I}_{nn'{\bf k}}+g^{\rm II}_{nn'{\bf k}}$ is used. Here, we set the momentum relaxation time parameter to $\eta=4$ meV. (b) Height of the orbital Hall conductivity plateau ($\bar{\sigma}_{\rm OH}$) as a function of $\Delta$, adopting the same color convention. The inset in panel (b) shows the evolution with $\Delta$ of the modulus of the difference between the heights of the plateaus with and without the inclusion of the $g^{\rm II}_{nn'{\bf k}}$ correction.}
	\label{FigOHE-2LGr} 
\end{figure}  

In contrast to the case of the bilayer of 2H-TMD, BLG has no orbital Hall conductivity within the intra-atomic approximation. This occurs because the low-energy spectrum of graphene is dominated by the orbitals $p_z$ of the carbon atoms. In Fig. \ref{FigOHE-2LGr}, we show the results for orbital Hall conductivity obtained by numerically evaluating Eq.(\ref{SigmaLR}) for the Hamiltonian of Eq. (\ref{HBGB}) using two values of $\Delta$. The blue curve shows the results considering only terms $m^{\rm SR}_{nn'{\bf k}}+g^{\rm I}_{nn'{\bf k}}$ [Eqs.(\ref{mSR}) and (\ref{gauge1})]. It is worth noting that the height of the orbital Hall conductivity plateau of BLG, when considering the matrix elements $m^{\rm SR}_{nn'{\bf k}}+g^{\rm I}_{nn'{\bf k}}$ between conduction and valence bands, increases by $\approx 30\%$ compared with previous results obtained within the semiclassical regime, where these elements were neglected \cite{Cysne-PhysRevResearch.6.023271}. The red curves show the results considering the complete description, $m^{\rm SR}_{nn'{\bf k}}+g^{\rm I}_{nn'{\bf k}}+g^{\rm II}_{nn'{\bf k}}$ [Eq. (\ref{mSR}-\ref{gauge2})]. Again, the effect of the new term $g^{\rm II}_{nn'{\bf k}}$ is to produce a decrease in the height of the OHE plateau. Nevertheless, in the case of bilayer graphene this decrease is smaller than the one obtained for bilayer 2H-TMD. 

In panel (b) of Fig. \ref{FigOHE-2LGr}, we show how the OHE plateau scales with the parameter $\Delta$ when the correction $g^{\rm II}_{nn'{\bf k}}$ is included (red curve) and when it is not (blue curve). For small values of $\Delta$ corresponding to small energy gaps in the electronic spectrum, the impact of the term $g^{\rm II}_{nn'{\bf k}}$ on the OHE plateau of BLG tends to increase. As $\Delta$ increases and consequently the energy gap in BLG becomes larger, the impact of $g^{\rm II}_{nn'{\bf k}}$ tends to decrease. It should be clear that this trend is not a general feature. The correction $g^{\rm II}_{nn'{\bf k}}$ depends on the complete electronic structure, including eigenvalues and eigenstates. Its role in the orbital Hall conductivity of materials must be analyzed on a case-by-case basis.

It should also be noted that when the Fermi energy lies inside a sufficiently large energy band gap, the orbital Hall conductivity plateau is robust against dilute disorder due to the absence of a Fermi surface. In this situation, our calculations based on pristine materials should fairly describe the OHE. Nevertheless, as the energy band gap becomes sufficiently small, disorder effects appear and may play a major role. The calculations presented here are not enough to represent the physical situation in this regime, and the inclusion of vertex corrections becomes necessary \cite{Liu2024, Tang-Bauer-Disorder, Veneri2025, Faridi2025}.

\section{conclusions}

In this work, we derived a formula for the orbital magnetic moment of Bloch states. In our derivation, we consistently include the Berry connection contribution in both the crystal momentum derivative of the Bloch state [Eq.(\ref{I1Ket})], and in the matrix element of the position operator [Eq.(\ref{rnnl})]. This inclusion, often neglected in the literature, is responsible for generating a new term in the expression for the orbital magnetic moment of Bloch states. The full expression [Eq.(\ref{OMMFull})] coincides with recent results obtained through a completely distinct approach, namely, the multipole expansion of the crystal’s optical matrix elements \cite{Urru2025}. We then applied the complete expression of the orbital magnetic moment to calculate the orbital Hall conductivity of the bilayer 2H-TMD and biased bilayer graphene. We find that the new term in the orbital magnetic moment expression is responsible for a reduction in the height of the orbital Hall insulating plateau in these materials. 

It is worth noting that the corrections $g^{\rm I}_{nn'{\bf k}}$ and $g^{\rm II}_{nn'{\bf k}}$ derived here appear only in matrix elements connecting Bloch states with different energies. We have verified their impact in single-layer models of graphene and transition-metal dichalcogenides and found that they do not affect the orbital Hall conductivity. We speculate that these corrections may have a stronger impact in bilayer systems with weak interlayer hopping. In such cases, the band structure contains non-degenerate Bloch states with relatively small energy separations, leading to a resonant enhancement of the correction terms.

Our derivation assumes non-degenerate energy bands. Extending it to include degenerate bands is an important step to be addressed in future work. In addition, a systematic study of the impact of the new contribution on orbital transport across different materials is also a relevant direction for future investigation in orbitronics.

To conclude, we recall that matrix elements $m^{\rm SR}_{nn'\mathbf{k}}+g^{\rm I}_{nn'\mathbf{k}}$ and $g^{\rm II}_{nn'\mathbf{k}}$ are separately gauge-covariant. They therefore provide natural building blocks for constructing gauge-invariant physical observables, such as response functions obtained after Brillouin-zone integration. This naturally raises the question of whether distinct gauge-invariant partial contributions to measurable quantities (for example, to the orbital Hall conductivity) can be defined starting from these building blocks. Similar conceptual developments occurred in the modern theory of equilibrium orbital magnetization: while early decompositions into local and itinerant circulation were not individually gauge invariant, a later regrouping produced two gauge-invariant contributions \cite{Ceresoli-2006}, and subsequent work related complementary probes (including integrated magnetic circular dichroism) to combinations of these gauge-invariant terms \cite{Ivo-PhysRevB.77.054438}. We further note that the Brillouin-zone integral of the correction $g^{\rm II}_{nn'\mathbf{k}}$ resembles (up to a factor of two) an interband analogue of the Berry-curvature contribution to the equilibrium orbital magnetization of a trivial insulator \cite{Urru2025,ModernTheory-PhysRevLett.95.137204}. Elucidating how these gauge-covariant contributions reorganize into physically measurable components of orbital transport is an interesting direction for future work.

\begin{acknowledgments}
We acknowledge CNPq/Brazil, CAPES/Brazil, FAPERJ/Brazil,  INCT Nanocarbono for financial support. TPC acknowledges CNPq (Grant No. 305647/2024-5).  TGR acknowledges  FCT - Fundação para a Ciência e Tecnologia,  project reference numbers UIDB/04650/2020, 2023.11755.PEX ( with DOI identifier https://doi.org/10.54499/2023.11755.PEX) and 2022.07471.CEECIND/CP1718/CT0001
(with DOI identifier: 10.54499/2022.07471.CEECIND/CP1718/
CT0001) and acknowledges support from the EIC Pathfinder OPEN grant 101129641 “OBELIX”. IS acknowledges Grant No.  PID2021-129035NB-I00 funded by
MCIN/AEI/10.13039/501100011033 and by ERDF/EU. We thank Michael Sch\"{u}ler for helpful discussions regarding Eq. (\ref{rnnl}).

\end{acknowledgments} 
 
\appendix
\section{Details on Eqs. (\ref{I1Ket}) and (\ref{rnnl}). \label{AppA}}
\subsection{Demonstrating the identity of Eq. (\ref{I1Ket})}

Considering that Hamiltonian $\hat{H}_{\bf k}$ has a non-degenerate energy spectrum, one can write the following relation:
\begin{eqnarray}
\braket{n|{\vec{\nabla}_{\bf k}m}}=\frac{\bra{n}\vec{\nabla}_{\bf k}\hat{H}_{\bf k}\ket{m}}{\epsilon_m-\epsilon_n}, \text{valid for } n\neq m.
\end{eqnarray}
Applying $\ket{n}$ and summing over $n\neq m$,
\begin{eqnarray}
\sum_{n\neq m}\ket{n}\frac{\bra{n}\vec{\nabla}_{\bf k}\hat{H}_{\bf k}\ket{m}}{\epsilon_m-\epsilon_n}&=&\left[\sum_{n\neq m} \ket{n}\bra{n}\right]\ket{\vec{\nabla}_{\bf k}m} \nonumber \\
&=&\left[\mathbb{1}-\ket{m}\bra{m}\right]\ket{\vec{\nabla}_{\bf k}m} \nonumber \\
&=& \ket{\vec{\nabla}_{\bf k}m}+i\vec{\mathcal{A}}_{m{\bf k}} \ket{m}. \nonumber \\ 
\end{eqnarray}
using that $\hat{\vec{v}}_{\bf k}=\hbar^{-1}\vec{\nabla}_{\bf k} \hat{H}_{\bf k}$ we obtain Eq. (\ref{I1Ket}). 

\subsection{Consistency of Eq. (\ref{rnnl}) with Eq. (\ref{I1Ket})}
Given the expression,
\begin{eqnarray}
\bra{n}\hat{\vec{r}}\ket{n'}=\bra{n}\left(i\hat{\vec{d}}-\mathbb{1}\vec{\mathcal{A}}_{n'{\bf k}}\right)\ket{n'}. \label{IDr}
\end{eqnarray}
Considering that position is a Hermitian operator, then $\hat{\vec{r}}=\hat{\vec{r}}^{\dagger}$, or
\begin{eqnarray}
i\hat{\vec{d}}-\mathbb{1}\vec{\mathcal{A}}_{n'{\bf k}}&=&\left(i\hat{\vec{d}}-\mathbb{1}\vec{\mathcal{A}}_{n'{\bf k}}\right)^{\dagger}\nonumber \\
&=&-i\hat{\vec{d}}^{\dagger}-\mathbb{1}\vec{\mathcal{A}}_{n'{\bf k}}
\end{eqnarray}
and we have $\hat{\vec{d}}=-\hat{\vec{d}}^{\dagger}$. Defining $\hat{\vec{d}}\ket{n}=\ket{\vec{\nabla}_{\bf k}n}$ we have $\bra{\vec{\nabla}_{\bf k}n}=\bra{n}\hat{\vec{d}}^{\dagger}=-\bra{n}\hat{\vec{d}}$. Then we obtain:
\begin{eqnarray}
\hat{\vec{d}}\ket{n}&=&\ket{\vec{\nabla}_{\bf k}n} \\
\bra{n}\hat{\vec{d}}&=&-\bra{\vec{\nabla}_{\bf k}n}
\end{eqnarray}
This gives the action of the operator $\hat{\vec{d}}$ in Hilbert space. 

Before we proceed, we demonstrate a useful relation. The velocity operator can be defined by $\frac{d\hat{\vec{r}}}{dt}=\hat{\vec{v}}=\frac{1}{i\hbar}\left[\hat{\vec{r}},\hat{H}\right]$. Given this,
\begin{eqnarray}
\bra{n}\hat{\vec{v}}\ket{n'}=\frac{1}{i\hbar}\bra{n}\left(\hat{\vec{r}}\hat{H}-\hat{H}\hat{\vec{r}}\right)\ket{n'}.
\end{eqnarray}
Then we obtain
\begin{eqnarray}
\bra{n}\hat{\vec{r}}\ket{n'}=\frac{i\hbar\vec{v}_{nn'}}{\epsilon_{n'}-\epsilon_{n}}.
\end{eqnarray}
This identity holds for $n\neq n'$, assuming that the Hamiltonian has a non-degenerate energy spectrum.

Now, starting from Eq. (\ref{IDr}),
\begin{eqnarray}
\bra{n}\hat{\vec{r}}\ket{n'}&=&i\braket{n|\vec{\nabla}_{\bf k}n'}-\braket{n|n'}\vec{\mathcal{A}}_{n'{\bf k}} \nonumber \\
&=&\frac{i\hbar\vec{v}_{nn'}}{\epsilon_{n'}-\epsilon_n}
\end{eqnarray}
We now multiply the above equation by $\ket{n}$ and sum over $n\neq n'$,
\begin{eqnarray}
\sum_{n\neq n'}\frac{i\hbar\vec{v}_{nn'}\ket{n}}{\epsilon_{n'}-\epsilon_n}&=&i\sum_{n\neq n'}\ket{n}\braket{n|\vec{\nabla}_{\bf k}n'}-\sum_{n\neq n'}\ket{n} \braket{n|n'}\vec{\mathcal{A}}_{n'{\bf k}} \nonumber \\
&=&\left(\sum_{n\neq n'}\ket{n}\bra{n} \right)\left(i\ket{\vec{\nabla}_{\bf k}n'}-\vec{\mathcal{A}}_{n'{\bf k}}\ket{n'}\right) \nonumber \\
&=&\left(\mathbb{1}-\ket{n'}\bra{n'}\right)\left(i\ket{\vec{\nabla}_{\bf k}n'}-\vec{\mathcal{A}}_{n'{\bf k}}\ket{n'}\right) \nonumber \\
&=& i\ket{\vec{\nabla}_{\bf k}n'}-\vec{\mathcal{A}}_{n'{\bf k}}\ket{n'}\nonumber \\
& &-i\ket{n'}\braket{n'|\vec{\nabla}_{\bf k}n'}+\vec{\mathcal{A}}_{n'{\bf k}}\ket{n'}\nonumber \\
&=&i\left(\ket{\vec{\nabla}_{\bf k}n'}+i\vec{\mathcal{A}}_{n'{\bf k}}\ket{n'}\right)
\end{eqnarray}
Swapping $n\leftrightarrow n'$ we obtain Eq. (\ref{I1Ket}). Then Eq. (\ref{IDr}) is consistent with Eq. (\ref{I1Ket}).

\section{Intermediate steps of Section \ref{SecmPhys} \label{APPmathDetails}}
Here, we show how $m_{nn'{\bf k}}$ can be expressed using the last equality of Eq. (\ref{mPhysIandII}) in the main text. Then, starting from the second line of Eq. (\ref{mPhysIandII}), we proceed as follows:
\begin{widetext}
\begin{eqnarray}
m_{nn'{\bf k}}&=&-\frac{e}{4}\sum_m\Big[-i\braket{\partial_x n|m}v^y_{mn'}+iv^y_{nm}\braket{m|\partial_xn'}-(x\leftrightarrow y)\Big]+\frac{e}{4}\Big[\mathcal{A}^x_{n{\bf k}}v^y_{nn'}+v^y_{nn'}\mathcal{A}^x_{n'{\bf k}}-(x\leftrightarrow y)\Big] \nonumber \\
&=& -\frac{e}{4}\sum_m \Bigg\{ \sum_{\tilde{m}\neq n}\frac{-i\hbar v^x_{n\tilde{m}}v^y_{mn'}\delta_{\tilde{m}m}}{\Delta\epsilon_{n\tilde{m}}}+\sum_{\tilde{m}\neq n'}\frac{i\hbar v^y_{nm}v^x_{\tilde{m}n'}\delta_{\tilde{m}m}}{\Delta\epsilon_{n'\tilde{m}}}-(x\leftrightarrow y)\Bigg\} \nonumber \\
&&-\frac{e}{4}\Big\{\mathcal{A}^x_{n{\bf k}}v^y_{nn'}+\mathcal{A}^x_{n'{\bf k}}v^y_{nn'}-\mathcal{A}^x_{n{\bf k}}v^y_{nn'}-\mathcal{A}^x_{n'{\bf k}}v^y_{nn'} -(x\leftrightarrow y) \Big\},
\end{eqnarray}
where, we used the identity from Eq. (\ref{I1Ket}). The terms proportional to Berry connections in the final part of the previous equation cancel out, and the remaining terms can be rearranged by taking into account those inside the parentheses $(x \leftrightarrow y)$:
\begin{eqnarray}
m_{nn'{\bf k}}&&= \frac{ie\hbar}{4}\sum_m \left\{ \sum_{\tilde{m}\neq n}\frac{v^x_{n\tilde{m}}v^y_{mn'}\delta_{\tilde{m}m}}{\Delta\epsilon_{n\tilde{m}}}+\sum_{\tilde{m}\neq n'}\frac{v^x_{nm}v^y_{\tilde{m}n'}\delta_{\tilde{m}m}}{\Delta\epsilon_{n'\tilde{m}}}-(x\leftrightarrow y)\right\} \nonumber \\
&&=\frac{ie\hbar}{4}\left(\Bigg\{ \sum_{\substack{m\neq n' \\ \tilde{m}\neq n}} \frac{v^x_{n\tilde{m}}v^y_{mn'}\delta_{\tilde{m}m}}{\Delta\epsilon_{n\tilde{m}}}+\sum_{\substack{m\neq n \\ \tilde{m}\neq n'}} \frac{v^x_{nm}v^y_{\tilde{m}n'}\delta_{\tilde{m}m}}{\Delta\epsilon_{n'\tilde{m}}}\Bigg\}_{i}+\Bigg\{ \sum_{\tilde{m}\neq n}\frac{v^x_{n\tilde{m}}v^y_{n'n'}\delta_{\tilde{m}n'}}{\Delta\epsilon_{n\tilde{m}}}+\sum_{\tilde{m}\neq n'}\frac{v^x_{nn}v^y_{\tilde{m}n'}\delta_{\tilde{m}n}}{\Delta\epsilon_{n'\tilde{m}}} \Bigg\}_{ii}\right)-(x\leftrightarrow y).\nonumber\\ \label{mPhysIandIIAPP}
\end{eqnarray}
In the second line, we used $\sum_mp(m)=p(n')+\sum_{m\neq n'}p(m)=p(n)+\sum_{m\neq n}p(m)$ to divide the sum over the index $m$. The last line corresponds to the expression used in the main text. After some algebra and dummy index swapping, one can identify the contributions from the first parentheses with the function $F_{nn'{\bf k}}$ defined in the main text [Eq. (\ref{Fnnp})]: $\frac{ie\hbar}{4}\big\{ ...\big\}_{i}-(x \leftrightarrow y)=F_{nn'{\bf k}}$. We also define a function that contains the contributions from Eq. (\ref{mPhysIandIIAPP}) enclosed in $\big\{...\big\}_{ii}$:
\begin{eqnarray}
\tilde{g}_{nn'{\bf k}}&&=\frac{ie\hbar}{4}\Bigg\{...\Bigg\}_{ii}-(x\leftrightarrow y) \nonumber \\
&&=\frac{e}{4}\left[v^y_{n'n'}i\braket{\partial_xn|n'}+v^x_{nn}i\braket{n|\partial_yn'}\right]\left(1-\delta_{nn'}\right)-(x\leftrightarrow y).
\label{mPhysII}
\end{eqnarray}
In the second equality, we used $\sum_{\tilde{m}\neq n} \bar{p}(\tilde{m}) \delta_{\tilde{m}n'}=\bar{p}(n')(1-\delta_{nn'})$, $\sum_{\tilde{m}\neq n'} \bar{p}(\tilde{m}) \delta_{\tilde{m}n}=\bar{p}(n)(1-\delta_{nn'})$, and the identity of Eq. (\ref{I1Ket}). We also discard the identically zero term $\big[v^y_{n'n'}\mathcal{A}^x_{n{\bf k}}-v^x_{nn}\mathcal{A}^y_{n'{\bf k}}\big]\delta_{nn'}\left(1-\delta_{nn'}\right)=0$. Through these passages, we demonstrated Eq. (\ref{mPhysIandII}) presented in the main text.
\end{widetext}

\section{Details on analytical calculations for OHE in bilayer of 2H-TMD \label{AppB}}

\subsection{First-Order perturbation theory in $t_{\perp}/\Delta$}

One can use degenerate perturbation theory to calculate the corrected energies and wave functions of Eq.(\ref{Heff}) to first order in interlayer coupling. This calculation is detailed in Ref. \cite{Cysne-Vignale-Tatiana-PhysRevB.105.195421} and yields the following results for the corrected energies:
\begin{eqnarray}
\tilde{E}_{v(c),\pm}(q)=E^0_{v(c)}(q)+\delta E_{v(c),\pm}(q), \label{CorrectedEnergies}
\end{eqnarray}
where 
\begin{eqnarray}
E^0_{v(c)}(q)=\frac{1}{2}\left(\Delta \mp \sqrt{\Delta^2+4a^2t^2q^2} \right), \label{E0}
\end{eqnarray}
is unperturbed energy dispersion associated to conduction ($c$) and valence ($v$) bands, and
\begin{eqnarray}
\delta E_{v,\pm}(q)=\pm \frac{t_{\perp}}{2} \left(1+\frac{\Delta}{\sqrt{\Delta^2+4a^2t^2q^2}}\right), \label{correctionVal} \\
\delta E_{c,\pm}(q)=\pm \frac{t_{\perp}}{2} \left(1-\frac{\Delta}{\sqrt{\Delta^2+4a^2t^2q^2}}\right). \label{correctionCond}
\end{eqnarray}
are corrections due to the interlayer hopping. The effect of interlayer coupling on the wave functions translates into the formation of bonding ($+$) and antibonding ($-$) linear combinations of the eigenstates of the individual layers in the valence and conduction subspaces:
\begin{eqnarray}
\ket{\Phi_{v(c),\pm}({\bf q}_{\tau})} =\frac{1}{\sqrt{2}} \left(\ket{\psi_{v(c),1}({\bf q}_{\tau})} \pm \ket{\psi_{v(c),2}({\bf q}_{\tau})} \right). \nonumber \\ 
\label{CorrectedStates}
\end{eqnarray}
where the unperturbed eigenvectors of the valence ($v$) band are
\begin{eqnarray}
\big|\psi_{v,1}({\bf q}_{\tau})\rangle= \mathcal{N}_v(q) \left(\frac{E^0_{v}(q)}{\gamma_-},1,0,0\right)^T, \label{PsiVal01} \\
\big|\psi_{v,2}({\bf q}_{\tau})\rangle= \mathcal{N}_v(q) \left(0,0, \frac{E^0_{v}(q)}{\gamma_+},1\right)^T, \label{PsiVal02} 
\end{eqnarray}
while the conduction ($c$) band eigenvectors are
\begin{eqnarray}
\big|\psi_{c,1}({\bf q}_{\tau})\rangle=\mathcal{N}_c(q) \left(\frac{E^0_c(q)}{\gamma_-},1,0,0\right)^T, \label{PsiCond01} \\
\big|\psi_{c,2}({\bf q}_{\tau})\rangle=\mathcal{N}_c(q) \left(0,0,\frac{E^0_c(q)}{\gamma_+},1\right)^T. \label{PsiCond02} 
\end{eqnarray}
The normalization factors are $\mathcal{N}_{v(c)}(q)=\left[1+(E^0_{v(c)}(q))^2/(a t q)^2 \right]^{-1/2}$ and the superscript $T$ means the application of transpose operation to obtain column vectors.

\subsection{Orbital Hall conductivity of bilayer 2H-TMD using the complete OMM expression}
\begin{widetext}
Using the results obtained in the main text [Eqs.(\ref{Lz})], we compute the orbital current flowing in the y-direction:
\begin{eqnarray}
J^{y,L_z}_{{\bf q}_{\tau}}=\frac{\tau a t}{\mu_{\rm B}} \left[
\begin{array}{cccc}
 a_q\sin (\theta) & -\frac{i}{2}m^0_q & 0 & 0 \\
 \frac{i}{2}m^0_q & a_q\sin (\theta) & 0 & 0 \\
 0 & 0 & -a_q\sin (\theta) & -\frac{i}{2}m^0_q \\
 0 & 0 & \frac{i}{2}m^0_q & -a_q\sin (\theta) \\
\end{array}
\right],
\label{Jy}
\end{eqnarray} 
where $a_q=(e/\hbar)(at/(4q))$. We now insert this operator, along with the energy and vector states given in Eqs (\ref{CorrectedEnergies}, \ref{CorrectedStates}), into the orbital Berry curvature. To obtain simplified expressions, we integrate over the angle $\theta$ [using $d^2{\bf q}=qdqd\theta$] and expand the curvature in powers of $t_{\perp}/\Delta$, retaining only the first correction. After a straightforward calculation, we obtain for the valence band,
\begin{eqnarray}
\int^{2\pi}_0\frac{d\theta}{(2\pi)} \Omega_{v,\pm,{\bf q}_{\tau}}^{L_z}=-\frac{e}{\hbar \mu_{\rm B}}\left[ -\frac{\Delta ^2 a^4t^4}{\left(\Delta ^2+4 q^2 a^2t^2\right)^{5/2}} \mp t_{\perp}\frac{a^4t^4\Delta^2 \left(\Delta +\sqrt{\Delta ^2+4 q^2 a^2t^2}\right)}{\left(\Delta ^2+4 q^2 a^2t^2\right)^{7/2}}\right] +\mathcal{O}\left(\frac{t^2_{\perp}}{\Delta^2}\right), \label{OmegaLzVal}
\end{eqnarray}
and for the conduction band,
\begin{eqnarray}
\int^{2\pi}_0\frac{d\theta}{(2\pi)} \Omega_{c,\pm,{\bf q}_{\tau}}^{L_z}=-\frac{e}{\hbar \mu_{\rm B}}\left[ \frac{\Delta ^2 a^4t^4}{\left(\Delta ^2+4 q^2 a^2t^2\right)^{5/2}}\mp t_{\perp}\frac{a^4t^4\Delta ^2\left(-\Delta+\sqrt{\Delta ^2+4 q^2 a^2t^2}\right)}{\left(\Delta ^2+4 q^2 a^2t^2\right)^{7/2}}\right] +\mathcal{O}\left(\frac{t^2_{\perp}}{\Delta^2}\right). \label{OmegaLzCond}
\end{eqnarray}
Note that the right-hand side of the above equations does not depend on $\tau$, so the contributions from both valleys to the orbital Berry curvature are equal. Now, we extrapolate the validity of the low-energy model of Eq. (\ref{Heff}) to short wavelengths [$q\rightarrow \infty$] and calculate the $q$-integral of curvatures of Eqs.(\ref{OmegaLzVal}, \ref{OmegaLzCond}), expressing the result in terms of Fermi momenta $\tilde{q}_{v(c),\pm}(E_{\rm F})$. The Fermi momentum is determined from the Fermi energy by solving the following equation $\tilde{E}_{v(c),\pm}(\tilde{q}_{v(c),\pm})=E_{\rm F}$ [see Eq. (\ref{CorrectedEnergies})]. Their analytical expressions can be found in Ref. \cite{Cysne-Vignale-Tatiana-PhysRevB.105.195421}, and their plot as a function of Fermi energy is shown in Fig. \ref{FigOHE-TMD} (a). After a straightforward calculation, we obtain the expressions for the orbital Hall conductivity,
\begin{eqnarray}
\sigma^{L_z}_{\rm OH}(E_{\rm F})=\sum_{\tau}\sum_{{\rm l.c.}=\pm}\left[\sigma^{L_z}_{v, {\rm l.c.}}(E_{\rm F})+\sigma^{L_z}_{c, {\rm l.c.}}(E_{\rm F})\right], \label{OHCfinal}
\end{eqnarray}
where the contribution from the bonding (${\rm l.c.}=+$) and antibonding (${\rm l.c.}=-$) states in the valence band is given by:
\begin{eqnarray}
\sigma^{L_z}_{v, \pm}(E_{\rm F})=g_s\left(\frac{e^2}{2\pi\hbar\mu_{\rm B}}\right)\left[ \frac{\Delta ^2 a^2t^2}{12 \left(\Delta ^2+4 a^2t^2\tilde{q}_{v,\pm}^2\right)^{3/2}}\pm t_{\perp}\frac{\Delta ^2 a^2t^2 \left(4 \Delta +5 \sqrt{\Delta ^2+4 a^2t^2\tilde{q}_{v,\pm}^2}\right)}{80 \left(\Delta ^2+4 a^2t^2\tilde{q}_{v,\pm}^2\right)^{5/2}}\right], \label{sigmaLzvFinal}
\end{eqnarray}
and, the contributions from the conduction band is
\begin{eqnarray}
&&\sigma^{L_z}_{c, \pm}(E_{\rm F})=g_s\left(\frac{e^2}{2\pi\hbar\mu_{\rm B}}\right)\Bigg[  -\frac{a^2t^2}{12\Delta}+\frac{a^2t^2\Delta^2}{12\left(\Delta^2+4 a^2t^2 \tilde{q}_{c,\pm}^2\right)^{3/2}} \nonumber \\
&&\hspace{55mm} \pm t_{\perp}\frac{a^2t^2 \left(\Delta ^5+\left(2 a^2t^2 \tilde{q}^2_{c,\pm}\Delta^2-\Delta^4+4 a^4t^4 \tilde{q}^4_{c,\pm}\right) \sqrt{\Delta ^2+4 a^2t^2 \tilde{q}^2_{c,\pm}}\right)}{20 \Delta ^2 \left(\Delta ^2+4  a^2t^2 \tilde{q}^2_{c,\pm}\right)^{5/2}}
\Bigg]. \label{sigmaLzcFinal}
\end{eqnarray}
We have included the spin degeneracy factor $g_s=2$ in the above expressions. Contrasting the expressions presented in the above equations, one notices that they differ from those reported in Ref. \cite{Cysne-Vignale-Tatiana-PhysRevB.105.195421} by a factor of 1/2. This difference can be traced back to the new term $g^{\rm II}_{nn'{\bf k}}$ obtained in this work.
\end{widetext}


%

\end{document}